\documentclass[10pt,twocolumn,prl,aps,amssymb,amsmath,showpacs,tightenlines]{revtex4}
\usepackage{graphicx}
\usepackage{todonotes}
\usepackage{framed,color}
\usepackage{tensor}
\usepackage{enumerate}
\usepackage{hyperref}

\newcommand{\half}{\mbox{$\textstyle \frac{1}{2}$}}

\begin{document}
\title{A jetlet hierarchy for ideal fluid dynamics}
\author{C. J. Cotter}
\author{D. D. Holm}
\author{H. O. Jacobs}
\author{D. M. Meier}
\date{\today}
\begin{abstract}
Truncated Taylor expansions of smooth flow maps are used in Hamilton's principle 
to derive a multiscale Lagrangian particle representation of ideal fluid dynamics. 
Numerical simulations for scattering of solutions at one level of truncation are 
found to produce solutions at higher levels. 
These scattering events to higher levels in the Taylor expansion are interpreted as modeling 
a cascade to smaller scales.
\end{abstract}
\pacs{numbers}
\maketitle


Particle-like solutions such as point vortices are an important tool in fluid dynamics \cite{Chorin1973}.
This paper discusses the dynamics of the multiscale hierarchy of local deformations of a fluid flow around any set of Lagrangian particles embedded in the flow introduced in \cite{JaRaDe2013, Sommer2013}.  This hierarchy carries the information found in a truncated Taylor series of the flow in the local regions around the chosen Lagrangian reference points. Truncating the Taylor series of a function at $k$-th order produces a polynomial that is called the $k$-jet of that function. We call each one of the embedded particles a `0-jetlet'. The hierarchy of multiscale information carried in the local deformations of the flow provided by truncating the Taylor expansion of the flow around one of the 0-jetlets at $k$-th order is called a `$k$-jetlet'. This paper derives the dynamical equations that govern the interactions among members of this multiscale hierarchy of local deformations around embedded particles in a fluid flow. The paper also provides revealing numerical simulations of these interactions. Its main results are:
\begin{enumerate}
\item We explicitly construct Hamiltonian dynamical equations for a hierarchy of jetlets in a regularized model of ideal incompressible fluid flow.
\item We find that these equations predict that jetlets carry their linear and angular momentum along a Lagrangian flow path while conserving their circulation.
\item In numerical simulations we investigate the particle-like behavior found in jetlet collisions and discover new types of nonlinear interactions among them.
\end{enumerate}

In deriving the equations of motion for this hierarchy of jetlets we use Lagrange multipliers to enforce a set of kinematic constraints that specify how the various terms in the expansion evolve.
We find that (a) the hierarchy is characterized by canonical Hamiltonian equations, (b) the solutions obey momentum conservation laws that can be understood as localized manifestations of Kelvin's circulation theorem, and (c) a pair of colliding jetlets converges asymptotically in time to a jetlet at a higher level in the hierarchy. We interpret (c) as a new representation of the well-known cascade to finer scales that occurs in nonlinear fluid flows \cite{Frisch1995}.
  \begin{figure}[t]
    \centering
    \includegraphics[width=0.45\textwidth]{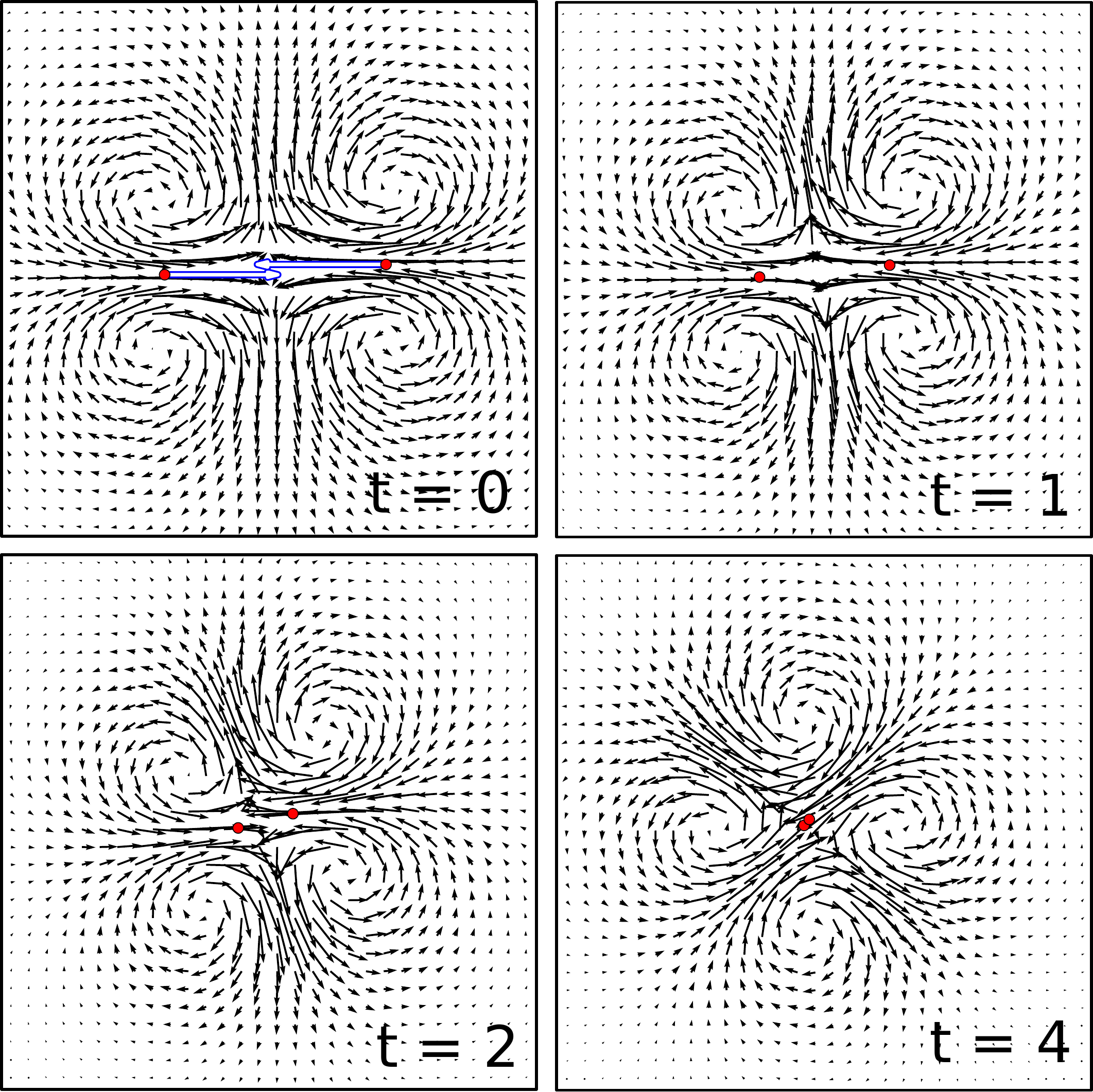}
    \caption{(color online) \emph{A jetlet collision.} The initial flow field is made up of two jetlets that belong to the lowest level in the hierarchy.  The jetlet locations are shown in red, while the arrows in the top left panel represent the initial linear momenta. At first, the jetlets propagate at constant speed; when they get close, they spiral in towards each other and combine to form a jetlet at the next higher level. The resulting structure propagates to the left and rotates, which is due to conservation of linear and angular momenta (see equation \eqref{eq:linear_and_angular_momenta} below).  A video is available online at \url{http://youtu.be/c1sRidMERWE}.  Additionally a concatenation of the previous video with a second video of single 1-jetlet depicts how a finite time merger could approximate the infinite time mergers given by the dynamics.  This video is available at \url{http://youtu.be/lUBAEyupcSM}. }

    
    \label{fig:collision}
  \end{figure}

The motion of a fluid in $n$ dimensions can be described by a time-dependent flow map $\varphi$, which specifies how each point $x$ in the reference configuration follows its path $q(t) =\varphi (x, t)$.
 Consequently $\dot{q}^i = u^i(q)$, where $u$ is the Eulerian flow field and $i = 1,\dots,n$ is a coordinate index.
 The time evolution of the flow field can be derived from Hamilton's principle, $\delta \int \ell(u)\, dt = 0$, where $\ell$ is the kinetic energy Lagrangian.
 For an ideal incompressible fluid we have $\ell(u) =  \half \|u\|_{L^2}^2 = \half \int |u|^2 \, d^nx$.
 However, in order to ensure sufficient regularity, we consider instead the regularized Lagrangian $\ell(u) = \half \|u\|_{H^k}^2 +\frac{1}{2\epsilon^2} \|{\rm div}u \|_{H^k}^2=\half \int u \cdot \mathcal{A} u \, d^nx$,
 where the momentum operator $\mathcal{A}$ is given by
\begin{align*}
  \mathcal{A} =  \left( 1 - \frac{\sigma^2}{k} \Delta
  \right)^k \left( 1 - \frac{1}{\epsilon^2} \nabla \circ {\rm div}\right).
\end{align*}
We will comment on the parameters $\sigma$ and $\epsilon$ shortly. As for $k$, note that the flow field belonging to a Dirac-delta distributed momentum is $k - (n+1)/2$ times continuously differentiable in space \cite{MuMi2013}. That is, if the momentum $m(x)  = \mathcal{A} u(x)$ is equal to $p\, \delta(x - y)$ for some constant vector $p$, then the flow field $u$ is in $C^{k - (n+1)/2}$. 

  Hamilton's principle for this Lagrangian leads to the EPDiff equation \cite{HoMa2004} governing the time evolution of the flow field 
\begin{equation}
\begin{aligned}
  &\partial_tm + u\cdot\nabla m + (\nabla u)^T \cdot m + m ({\rm div}u) = 0   \\
  &u = K * m, \label{eq:EPDiff}
\end{aligned}
\end{equation}
where $*$ denotes convolution and $K$ is the Green's function for the operator $\mathcal{A}$.
Recently, it was shown that smooth solutions of these equations exist for all time so long as $\sigma > 0$ and $k$ is large enough.
Moreover, these solutions converge to solutions of the incompressible Euler equation when both $\epsilon$ and $\sigma$ tend to zero, with error bounds proportional to $\epsilon$ and $\sigma^2$ \cite{MuMi2013}.
Some other interesting choices of parameters are the following: For $\epsilon = \infty$ and $k = 1$ one obtains the $n$-dimensional Camassa--Holm equation \cite{CamassaHolm1993}; for $\epsilon = 0$ and $k = 1$ one obtains the Euler-$\alpha$ model for incompressible fluids \cite{HoMaRa1998}. 

In the present paper our main interest lies with PDEs that approximate ideal incompressible fluid dynamics.
Hence we consider the limit in which $k \to \infty$ and $\epsilon \to 0$.
In this limit the (matrix-valued) Green's function for $\mathcal{A}$ in two spatial dimensions becomes 
\begin{equation}
\begin{aligned}
  K^{ij}&(x) =   \frac{1}{4\pi\sigma^2} \left(e^{-\frac{r^2}{4 \sigma^2}} - \frac{2 \sigma^2}{r^2}\left(1 - e^{-\frac{r^2}{4\sigma^2}}\right)\right) \delta^{ij} \\
&+  \frac{1}{4\pi\sigma^2} \left(\frac{4 \sigma^2}{r^2} \left(1 - e^{-\frac{r^2}{4\sigma^2}}\right) - e^{-\frac{r^2}{4\sigma^2}}\right)\frac{x^ix^j}{r^2}, \label{eq:Greens}
\end{aligned}
\end{equation}
where $r = \|x\|$, and similar expressions can be derived for higher dimensions \cite{MuMi2013}. By appropriate rescalings of space and time we may set $\sigma = 1$ as a proxy for any $\sigma > 0$.
The Green's function \eqref{eq:Greens} was recently proposed in \cite{MiGl2014} for incompressible diffeomorphic image registration.

Having specified the fluid model, we proceed to construct an infinite hierarchy of jetlets that solve \eqref{eq:EPDiff}.
Our approach involves spatial Taylor expansions of the flow map $\varphi$.
The spatial derivatives of $\varphi$ are called `deformation gradients' in the theory of nonlinear elasticity \cite{MaHu1993}, a convention we follow here.
The first two deformation gradients are thus given by  $Q^i_j = \partial \varphi^i/\partial x^j$ and
$\mathcal{Q}^i_{jk} = \partial^2 \varphi^i/\partial x^j \partial x^k$.
As the flow map evolves in time, these quantities evolve correspondingly.
Namely, the time derivatives of the first deformation gradients evaluated at a point $x$ in the reference configuration are
\begin{align*}
\begin{aligned}
  &\dot{Q}^i_j = u^i_{,k}(q) Q^k_j \quad \mbox{and} \\
  &\dot{\mathcal{Q}}^i_{jk} = u^i_{,r\ell}(q) Q^r_j Q^\ell_k + u^i_{,\ell}(q) \mathcal{Q}^\ell_{jk},
\end{aligned}
\end{align*}
where sums are implied over repeated indices and we recall that $q(t) = \varphi(x, t)$.
Similar expressions can be derived for higher order deformation gradients.
We enforce these relations as constraints in Hamilton's principle by introducing Lagrange multipliers.
That is, we define the constrained action integral
\begin{align*}
  &S =
  \int \big[ \ell(u) 
  +  p_i (\dot{q}^i-u^i(q))
  +  P_i^j\big(\dot{Q}^i_j - u^i_{,r}(q)  Q^r_j\big)\notag \\ 
  &\,\,+ \mathcal{P}_i^{jk} \big( \dot{\mathcal{Q}}^i_{jk} - u^i_{,r\ell}(q) Q^r_j Q^\ell_k + u^i_{,\ell}(q) \mathcal{Q}^\ell_{jk} \big) + \ldots\big] \,  dt
\end{align*}
and require $\delta S = 0$ with respect to arbitrary variations of $u$, $q$, $p$, $Q$, $P$, $\mathcal{Q}$, $\mathcal{P}$, etc.
Whenever we keep track of several points $x_1$, $\cdots$, $x_N$  in the reference configuration, we will index them using Greek indices and write $Q^i_{\alpha j}$ and $\mathcal{Q}^i_{\alpha jk}$ for the deformation gradients at $x_\alpha$.
The corresponding Lagrange multipliers will be denoted by  $P^j_{\alpha i}$ and $\mathcal{P}^{jk}_{\alpha i}$. Sums over Greek indices will be stated explicitly.

\begin{figure}
    \centering
    \includegraphics[width=0.45\textwidth]{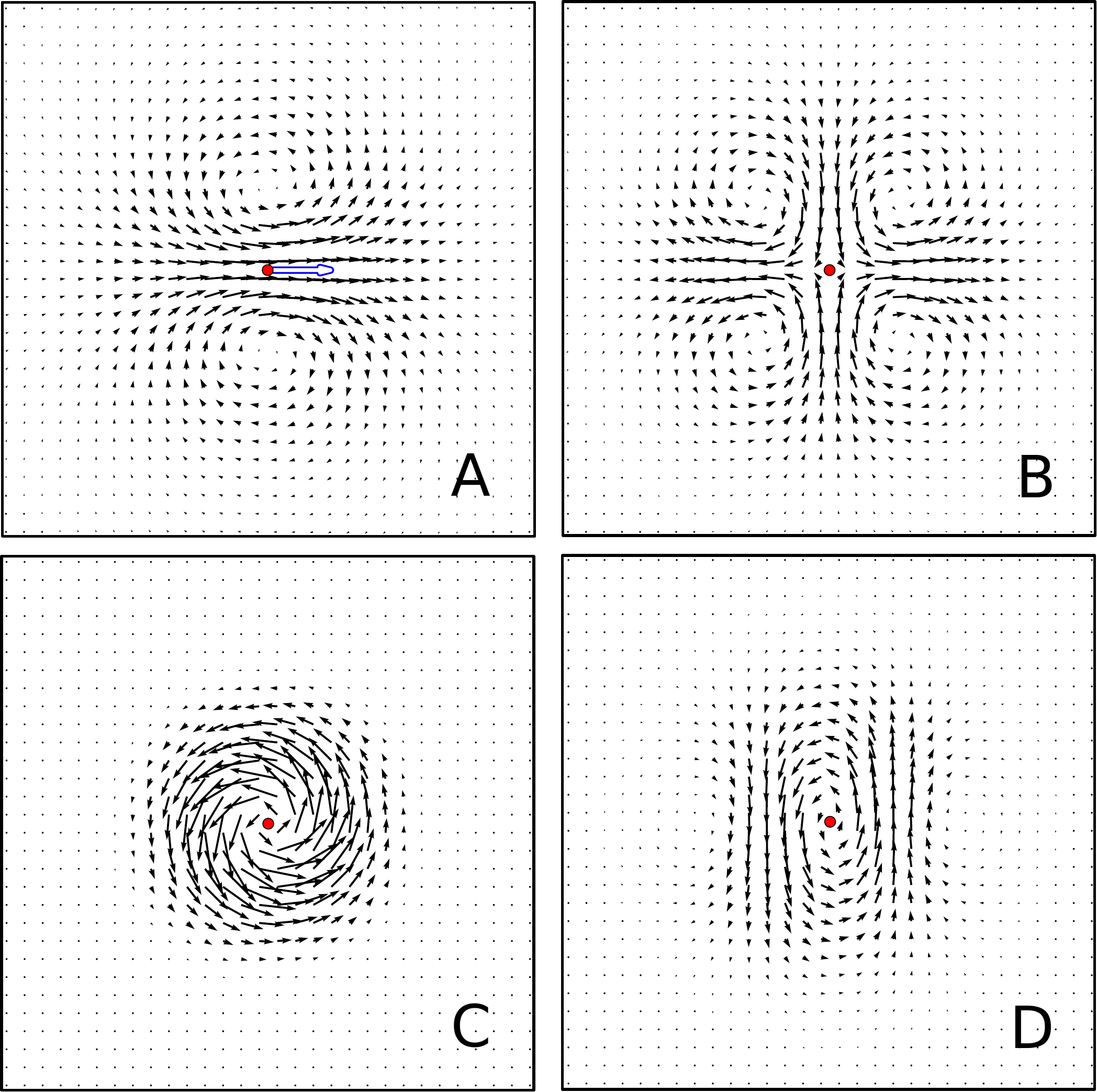}
    \caption{(color online) {\emph{First and second levels of the jetlet hierarchy}.} 
      Panel A depicts a velocity field obtained from \eqref{eq:0jet_u} with $p=(1,0)$.
      A video of the resulting dynamics for this figure is available online at \url{http://youtu.be/7jmZAnXg-lc}.
      Panels B, C, and D depict velocity fields obtained from \eqref{eq:1jet_u} with $p=(0,0)$ and matrices $\mu_B,\mu_C,\mu_D$ given in equation \eqref{eq:zoo}.
      The video associated to panel D is available online at \url{http://youtu.be/kB9gz9aPIa8}.
    }
    \label{fig:zoo}
  \end{figure}

The action integral above can be truncated so as to involve deformation gradients up to a certain order only. Each such truncation corresponds to a level in the hierarchy of jetlets. For instance, truncating at zeroth order we obtain
\begin{align}
   S_0 = \int \big[ \ell(u)   + \sum_{\alpha} p_{\alpha i} (\dot{q}_\alpha^i-u^i(q_\alpha)) \big] \,  dt, \notag
\end{align}
and the requirement $\delta S_0 = 0$ leads to the following equations of motion,
  \begin{align}
    &\dot{q}_\alpha^j = u^j(q_\alpha), \quad 
    \dot{p}_{\alpha j} = - u^k_{,j}(q_{\alpha}) p_{\alpha k}, \label{eq:q_dot,p_dot} \\
    &\mathcal{A}u(x) = \sum_{\alpha} p_\alpha \delta(x - q_\alpha). \label{eq:0jet_momentum}
  \end{align}
The third equation implies that
\begin{align}
  u^j(x) = \sum_{\alpha} K^{jk}(x - q_{\alpha}) p_{\alpha k}. \label{eq:0jet_u}
\end{align}
We can substitute the right hand side of \eqref{eq:0jet_u} into \eqref{eq:q_dot,p_dot} to obtain a finite dimensional ODE in the variables $q_\alpha$ and $p_\alpha$.
It is straightforward to verify that this ODE can be written as a set of canonical Hamiltonian equations with the kinetic energy Hamiltonian
\begin{align*}
  H = \frac{1}{2} \sum_{\alpha, \beta} p_{\alpha i} K^{ij}(q_\alpha - q_\beta) p_{\beta j}.
\end{align*}
 Upon integrating these equations to obtain $q_\alpha(t)$ and $p_\alpha(t)$, one can invoke \eqref{eq:0jet_u} to obtain a time-dependent velocity field, $u$, which solves \eqref{eq:EPDiff}, see \cite{HoMa2004}.
 In other words, we have found a class of solutions to \eqref{eq:EPDiff} parametrized by a finite-dimensional subspace of initial conditions.
 We call these solutions  0-jetlets and interpret $q_\alpha(t)$ and $p_\alpha(t)$ as the positions and momenta of the 0-jetlet, respectively.
  For two spatial dimensions these particles are called `vortons' in \cite{MuMi2013} and have the dipole-like structure that can be seen in panel A of Fig. \ref{fig:zoo}.
The width parameter, $\sigma$,  provides a spatial resolution to the velocity fields in \eqref{eq:0jet_u}.
Single jetlet solutions of this type can therefore be considered to model fluid velocity at scales greater than $\sigma$. 

To get the next level of the jetlet hierarchy 
we include deformation gradients up to first order in the action integral, which then becomes
\begin{align}
  S_1 &= 
  \int \big[ \ell(u) 
  + \sum_\alpha  p_{\alpha i} (\dot{q}_\alpha^i-u^i(q_\alpha)) \notag
  \\ &\quad +  P_{\alpha i}^j\big(\dot{Q}^i_{\alpha j} - u^i_{,r}(q_\alpha)  Q^r_{\alpha j}\big) 
 \big] dt. \notag
\end{align}
A calculation shows that Hamilton's principle implies 
\begin{align}
(\mathcal{A}u)_i(x) = \sum_\alpha p_{\alpha i} \delta(x - q_\alpha) - P^j_{\alpha i} Q^r_{\alpha j} \,\delta_{,r}(x - q_\alpha), \label{eq:Du1_onejet}
\end{align}
where $\delta_{,r}(x-q_\alpha)$ is the distributional derivative of the Dirac delta function. That is, for any test function $\phi$ we have
\begin{align}
  \int \phi(x) \delta_{,r}(x - q_{\alpha}) \,d^nx = -\phi_{,r}(q_\alpha). \notag
\end{align}
Using the Green's function $K$ from \eqref{eq:Greens} we conclude that
\begin{align}
  \! u^j(x) =  \sum_{\alpha} K^{jk}(x - q_{\alpha}) p_{\alpha k} - K^{jk}_{,r}(x - q_\alpha) Q^r_{\alpha \ell} P^\ell_{\alpha k}. \label{eq:1jet_u}
\end{align}
As concerns the equations of motion for $q$, $p$, $Q$ and $P$ one can verify that they are a set of canonical Hamiltonian equations with the kinetic energy Hamiltonian
\begin{align}
H &= \frac{1}{2} \sum_{\alpha, \beta} p_{\alpha i}K^{ij}(q_\alpha - q_\beta) p_{\beta j} +  2 p_{\beta j} P_{\alpha i}^r \notag
    Q_{\alpha r}^k  K_{,k}^{ij}(q_\alpha -q_\beta)  \\ &\qquad -  P_{\alpha i}^r Q_{\alpha r}^n P_{\beta j}^s Q_{\beta s}^k K^{ij}_{,nk}(q_\alpha - q_\beta). \notag
  \end{align}
Note that this Hamiltonian is invariant under the variable transformation
\begin{align}
  Q^i_{\alpha j} \to Q^i_{\alpha k} g^k_{\alpha j}, \qquad P^i_{\alpha j} \to (g_\alpha^{-1})^i_k P^k_{\alpha j} \label{eq:symmetry}
\end{align}
for any set of $n \times n$ matrices $g_1$, $\cdots$, $g_N$ of unit determinant.
This means that the equations of motion can be written in terms of the reduced variables $(q_\alpha, p_\alpha, \mu_\alpha)$, where
\begin{align*}
  \mu^i_{\alpha j} = Q^i_{\alpha k} P^k_{\alpha j}.
\end{align*}
One finds that
\begin{align*}
&\dot{q}_\alpha^i = u^i(q_\alpha), \quad \dot{p}_{\alpha i} = -u^j_{,i}(q_{\alpha}) p_{\alpha j} - u^j_{,ki}(q_\alpha)  \mu_{\alpha j}^k \\
&\dot{\mu}_{\alpha i}^j = u^j_{,k}(q_\alpha)\mu^k_{\alpha i} + \mu^j_{\alpha k} u^k_{,i}(q_\alpha).
 \end{align*}
By Noether's theorem the quantities
\begin{align}
  M_{\alpha i}^j = (Q_\alpha^{-1})^j_\ell \mu_{\alpha k}^\ell  Q_{\alpha i}^k \label{eq:Kelvin}
\end{align}
are constant in time.

 The symmetry transformation in \eqref{eq:symmetry} is a localized manifestation of the particle relabeling symmetry of the fluid.
 This can be seen as follows.
 If the flow map $\varphi$ is multiplied from the right by a volume preserving diffeomorphism $\psi$ that leaves the points $x_\alpha$ invariant, then the first deformation gradients transform as
\begin{align}
  Q^i_{\alpha j} = \frac{\partial \varphi^i}{\partial x^j}(x_\alpha) \to \frac{\partial \varphi^i}{\partial x^k}(x_\alpha) \frac{\partial \psi^k}{\partial{x^j}}(x_\alpha) =  Q^i_{\alpha k} \frac{\partial \psi^k}{\partial{x^j}}(x_\alpha) \notag.
\end{align}
These transformations are indeed of the form given in \eqref{eq:symmetry}. The conservation law associated with the particle relabeling symmetry is the circulation of the flow field around a closed loop advected by the fluid. Hence, the conservation of the quantities in \eqref{eq:Kelvin} is a localized version of Kelvin's circulation theorem, a fact previously noted in \cite{JaRaDe2013}.

  In addition to the previous symmetry, there are also translation and rotation symmetries, which yield the conserved linear and  angular momenta
\begin{align}  
  \begin{aligned} J_i &= \sum_\alpha p_{\alpha i}, \\
  \mathbf{J}_i^j &=\frac{1}{2} \sum_\alpha q_\alpha^j p_{\alpha i}  - q_\alpha^i p_{\alpha j} +   \mu^j_{\alpha i}  -  \mu^i_{\alpha j}.
  \end{aligned} \label{eq:linear_and_angular_momenta} 
\end{align}
Panels B,C, and D of Fig. \ref{fig:zoo} illustrate single jetlet flow fields of the form \eqref{eq:1jet_u} with $p = 0$ and
\begin{align}
  \mu_{B,C,D} = 
  \left(\begin{array}{cc} 1 & 0 \\ 0 & -1 \end{array}\right)  , \,  
  \left(\begin{array}{cc} 0 & -1 \\ 1 & 0 \end{array}\right) \mbox{ and } 
  \left(\begin{array}{cc} 0 & 1 \\ 0 & 0 \end{array}\right). \label{eq:zoo}
\end{align}
    The first two choices $\mu_B$ and $\mu_C$ yield stationary solutions of \eqref{eq:EPDiff}, while for the third choice $\mu_D$ the flow field evolves by rotations about the jetlet location.  Angular momentum is non-zero for the second and third cases. 

 Before we comment on the higher levels in the jetlet hierarchy, let us show what happens during jetlet collisions.
 A solution of the form \eqref{eq:0jet_momentum} that contains two jetlets satisfies
\begin{align}
  \mathcal{A}u(x) = p_1 \delta(x - q_1)  + p_2 \delta(x - q_2), \notag
\end{align}
whose integral against a vector valued test function $\phi$ is
\begin{align}
  \int \phi(x) \cdot \mathcal{A}u(x) \, d^nx = p_1 \cdot \phi(q_1) + p_2 \cdot \phi(q_2). \label{eq:test_f}
\end{align}
We define the difference vector $\delta q = (q_2 - q_1)/2$ and set $q = (q_1 + q_2)/2$, so that $q_1 = q - \delta q$ and $q_2 = q + \delta q$. If $q_1$ and $q_2$ are close to one another, we can approximate the right hand side of \eqref{eq:test_f} by
\begin{align}
  (p_{1 i} + p_{2 i}) \phi^i(q) + (p_{2 i} - p_{1 i}) \phi^i_{,k}\delta q^k + \mathcal{O}(\delta q^2). \notag
\end{align}
A comparison with \eqref{eq:Du1_onejet} suggests that if $\delta q$ tends to zero asymptotically,  these \emph{two} jetlets merge into a \emph{single} jetlet at the next higher level in the hierarchy with
\begin{align}
  p_i &= \lim_{t \to \infty} \,[p_{1 i} + p_{2 i}] \quad \mbox{and} \notag \\
\mu_i^j &=  \lim_{t \to \infty} \, [(p_{2 i} - p_{1i})\delta q^j]. \notag
\end{align}
Such mergers are consistent with angular momentum conservation, as one can verify using \eqref{eq:linear_and_angular_momenta}.

Numerical experiments demonstrate that these collisions indeed occur.  
An example of such an event is shown in Fig. \ref{fig:collision}: As the jet lets approach one another, the difference between the flow fields from the two $0$-jetlet solution and the corresponding single $1$-jetlet solution at the next higher level falls below machine precision. 
This behavior can be interpreted as the first step in the cascade of energy to smaller scales. 


To elaborate on this interpretation, we recall the definition of 
a Taylor expansion of a function $f$,
 \begin{align*}
   f(x) = f(0) + x f'(0) + \frac{1}{2} x^2 f''(0) + o(x^2)\,.
 \end{align*}
 The truncated expansion holds when $x$ is small compared to a length scale $\sigma$ provided the derivatives of $f(x/\sigma)$ are of order $O(1)$.
 Thus, we obtain a notion of scale for a given function $f$ based on its derivatives.
 In particular, a function $f$ which is described accurately in some neighborhood by a first-order Taylor expansion 
depends on finer scales than a function which is described sufficiently well by its zeroth-order Taylor expansion (i.e. a constant function).
When the fluid velocity $u$ is described well by its zeroth-order Taylor expansion, collision events will yield new velocity fields which are \emph{not} described well by zeroth-order Taylor expansions, but instead are well-described by first-order Taylor expansions.  From this perspective, the  collision behavior observed here describes a cascade towards finer scales.  


In this paper numerical simulations of the particle-like behavior found in jetlet collisions have revealed new types of nonlinear interactions among jetlets that produce higher-level jetlets from the collisions of lower level jetlets.
The equations of motion for the first levels in the jetlet hierarchy can be extended in a straightforward manner to the higher levels. The resulting equations of motion have the same character as the ones described above.
Namely, they are finite-dimensional ordinary differential equations in canonical Hamiltonian form that satisfy a localized version of Kelvin's circulation theorem. 
Moreover, when two jetlets of the same level collide, they combine to form a single jetlet at the next higher level in the hierarchy.  

We thank Dorje Brody, Martins Bruveris, Jaap Eldering, Sarang Joshi, and Stefan Sommer for helpful remarks and clarifying conversations. DDH, HOJ and DMM are grateful for partial support by the European Research Council Advanced Grant 267382 FCCA.

\end{document}